\documentclass[aps,prl,twocolumn,showpacs]{revtex4}
\usepackage{graphicx}

\begin{document}

\title{Forbidden island heights in stress-driven coherent Stranski-Krastanov
growth}

\author{Jos\'e Emilio Prieto}
\email{joseemilio.prieto@uam.es}
\affiliation{Centro de Microan\'alisis de Materiales and Instituto
Universitario ``Nicol\'as Cabrera'', Universidad Aut\'onoma de Madrid, E-28049
Madrid, Spain}

\author{Ivan Markov}
\email{imarkov@ipchp.ipc.bas.bg}
\affiliation{Institute of Physical Chemistry, Bulgarian Academy of Sciences,
1113 Sofia, Bulgaria}
\date{\today}

\begin{abstract}

%The observed height distribution of clusters obtained in strained epitaxy, 
%in particular the absence of certain heights, has been often interpreted
%in terms of electronic effects. Here we show that some aspects
%can be explained classically by the interplay of strain and edge energies.
%We find that soft materials grown on stiffer substrates 
%can transform directly from monolayer into thicker is\-lands by
%two-dimensional (2D) multilayer nucleation and growth. A critical thickness 
%exists which is a decreasing function of the force constant of the material's
%bonds. Thinner islands of soft materials are thermodynamically forbidden,
%due to the insufficient stress relaxation upon clustering particularly when
%they are under tensile stress. At sufficiently large misfits the barrier for
%2D mul\-ti\-layer nucleation is significantly smaller than the barrier for 
%subsequent single-layer nucleation. The effects are found to be
%quantitatively reasonable and offer a plausible explanation for the 
%absence of thin islands and 2D growth of flat-top islands usually 
%attributed to quantum size effects.

The observed height distribution of clusters obtained in strained epitaxy 
has been often interpreted in terms of electronic effects. 
We show that some aspects can be explained classically by the 
interplay of strain and edge energies. We find that soft materials 
can transform directly from monolayer into thicker islands by
two-dimensional (2D) multilayer nucleation and growth. There is a critical 
thickness decreasing with the force constant.
Thinner islands are thermodynamically forbidden, due to the insufficient 
stress relaxation upon clustering particularly under tensile stress. 
At sufficiently large misfits the barrier for 2D multilayer nucleation 
is significantly smaller than the barrier for subsequent single-layer 
nucleation. The effects are found to be quantitatively reasonable and 
offer a plausible explanation for the absence of thin islands and 
2D growth of flattop islands usually attributed to quantum size effects.

\end{abstract}

\pacs{68.35.Md, 68.43.Hn, 68.55.Ac, 68.65.Hb}

\maketitle

Nanostructures are very promising for optoelectronic and magnetic 
applications. For efficient operation, the shape, size and thickness 
distribution of small clusters are impor\-tant parameters. Therefore it 
is crucial to understand the factors which control them. 
The epitaxy of metals on 
se\-mi\-con\-ductor surfaces at low temperatures (130K - 180K) has been 
intensively studied in the last years~\cite{Mike1,Mike2,TTT1,Floreano}.
Some important observations are:  (i)
flat-top Pb islands with steep edges and a preferred height of 7 monolayers
(ML) grow on the wet\-ting lay\-er on
Si(111)$7\times 7$~\cite{Mike1,Mike2,TTT1,Yeh}, (ii) islands with thicknesses
from 1 to 3 MLs are never observed~\cite{TTT1,TTT2,Ozer}; (iii) flat-top
Pb islands are preceded by pyramidal or dome-like clusters; clus\-ters
thinner than 3 MLs are never registered~\cite{TTT3}; (iv) 2-ML thick 
flat-top Ag islands on Si(111) increase linearly in size with increasing 
coverage whereas ML islands
preserve a nearly constant size of about 5 nm$^2$~\cite{TTT3}; (v) flat-top
islands with a preferred height grow laterally without 
thickening~\cite{Mike1,Ozer,Budde}; (vi) Vertical growth of Pb/Si(111) takes
place by bilayer increments~\cite{Hupalo,Ozer}.

The above observations were explained in terms of the energy lowering 
due to electron confinement and spilling of charge through the
metal-semiconductor interface by Zhang, Niu and Shih, who coined for this 
reason the term ``electronic growth"~\cite{Zhenyu}. However, classical 
effects associated to strain relaxation are expected to contribute, as 
assumed to explain the fact that the increase of the aspect ratio 
of flat-top Pb islands is mediated by the formation and growth of strips 
(or rings) around the outer edges of the islands~\cite{Okam}.
The aim of this paper is to show that some of the observations listed above,
in particular (ii), (iii), (iv) and (v), can be explained classically in 
terms of the interplay of strain and edge energies.

Two mechanisms have been invoked to address the in\-sta\-bi\-li\-ty of planar
growth against clustering. The first is the nucleationless development of
instabilities of a certain wavelength that evolve into faceted 
three-dimensional (3D)
islands~\cite{Sutter}. The second is a nucleation behavior due to the
competition of surface energy and strain relaxation~\cite{Tersoff}.
Studies of the total energy per atom of islands of different heights
as a function of their total number of atoms have shown that intermediate
states with thick\-nesses increasing by ML steps are stable in separate
con\-se\-cutive intervals of volume~\cite{Khor,Pri1}. 
The rearrangement of mono- to bilayer islands was found to be a
true nucleation pro\-cess in compressed overlayers, in the sense that a small
critical nucleus of the second layer is initially formed and then grows
further up to the completion of the transformation. The associated energy
displays a maximum at some number of atoms in the second level, which 
becomes very small for high enough values of the lattice misfit~\cite{Pri2}.
However, bilayer islands of expanded overlayers tend to require 
unrealistically high values of the
lattice strain and of the is\-land size to become stable against ML
islands. Their trans\-formation curves show an increase of the energy up to
large second layer cluster sizes, making nucleation un\-like\-ly~\cite{Pri2}. 
Also compressed overlayers of softer materials, or at lower misfits,
are not expected to show nucleation behavior through this mechanism. 
We therefore set out to study the mono- to multilayer island transformation 
by using a simple atomistic model in $2+1$ dimensions.

We consider fcc(100) crystallites with the shape of truncated
squ\-are py\-ra\-mids on a rigid layer of the same material
with lattice constant $a$. 
In our model, atoms interact through a pair potential 
\begin{equation}\label{potent}
V(r) = V_{0}\Biggl[\frac{\nu }{\mu - \nu }e^{-\mu (r-b)} - \frac{\mu }{\mu -
\nu }e^{-\nu (r-b)}\Biggr]. 
\end{equation}
that in spite of its simplicity includes all necessary features to describe
real materials (its strenght and an\-har\-mo\-ni\-ci\-ty are governed by 
the constants $\mu $ and $\nu $~\cite{Mar2}).
The equilibrium atom separation is $b$, so that the lattice misfit is 
given by $\varepsilon = (b-a)/a$.  
We consider interactions only in the first coordination sphere. 
For the study of mono- to multilayer trans\-for\-ma\-tions, we
assume model processes in which atoms are detached one by one
from the edges of an initial square ML island, placed on top of it 
at the center and arranged in a $(X-1)$-ML thick structure as compact as 
possible up to the formation of a $X$-ML truncated pyramid.
For every intermediate configuration, the cluster is allowed to
relax by varying iteratively the atomic positions until all the forces 
fall below some negligible cutoff value. 
The energy change associated with 
the transformation at a particular stage is given by the difference between 
the energy of the incomplete multilayer island and that of the initial 
ML island.

\begin{figure}[ht]
\includegraphics*[width=5.5cm]{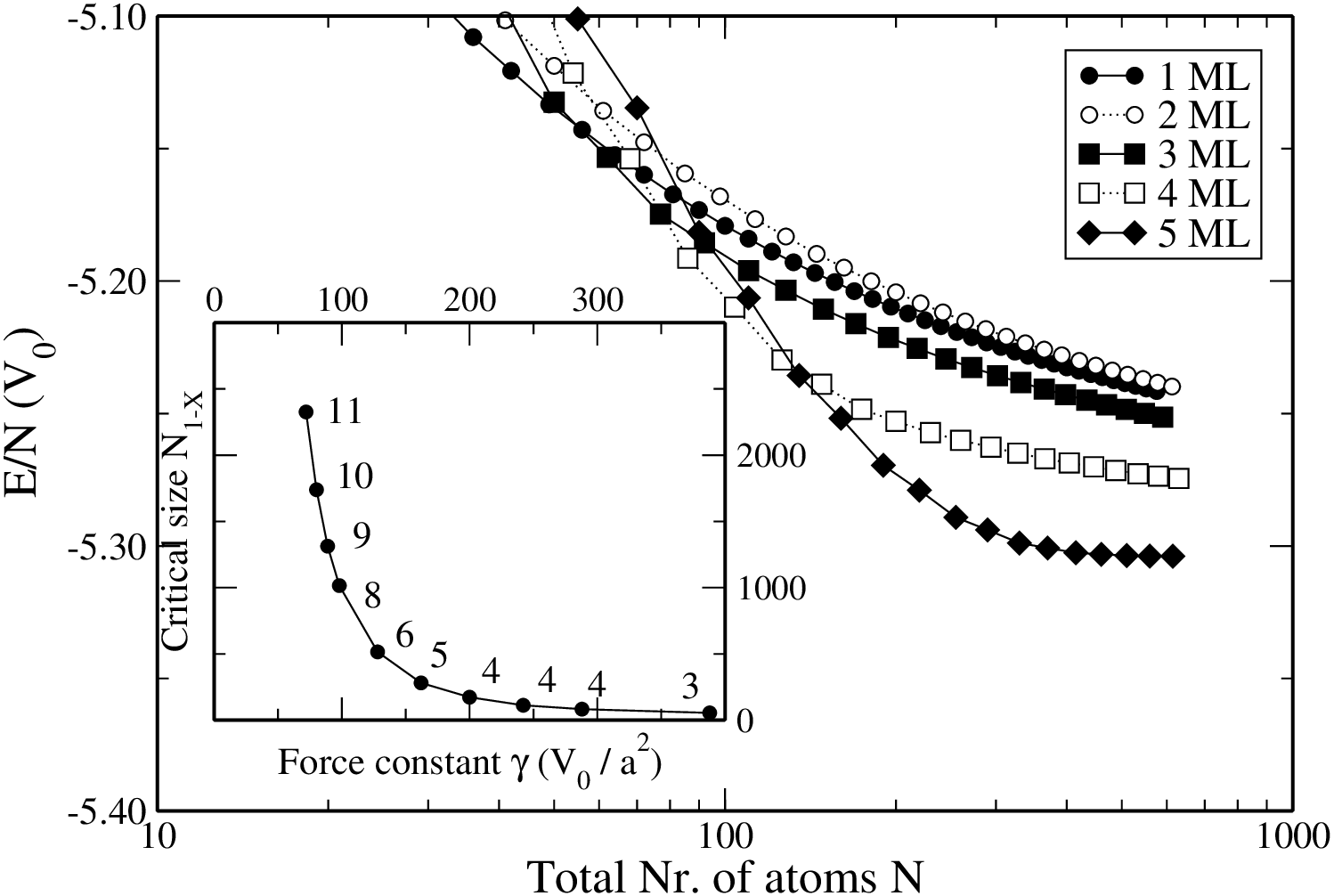}
\caption{\label{EN2613-07} Energy per atom of mono- and multilayer
islands as a function of the total number of atoms. The misfit $\varepsilon$
amounts to -7\% and $\mu = 2\nu = 26$.
The insert shows the dependence of the critical size $N_{1-X}$ (beyond 
which 1-ML islands become unstable against $X$-ML islands) on the force 
constant $\gamma $. The threshold thickness $X$ is denoted by the numbers 
at each point.}
\end{figure}

Figure~\ref{EN2613-07} shows the total energy per atom of expanded islands 
of different thicknesses as a function of the total number of atoms. 
It is seen that monolayer islands remain always stable against bi\-layer ones. 
At a critical size $N_{1-3}$, tri\-layer islands be\-come stable and, with 
sizes increasing further, the ground state shifts to islands of 
thicknesses increasing in ML-steps.
The insert of Fig.~\ref{EN2613-07} illustrates the dependence of the 
critical size $N_{1-X}$ on the force constant $\gamma $ ($\mu \nu V_0$ in our
model) for a fixed value of the misfit ($\varepsilon$=-0.07). 
The threshold thickness below which
multilayer islands are energetically unfavored steep\-ly increases together
with the critical volume with in\-creas\-ing softness (decreasing $\gamma $)
of the epitaxial over\-lay\-er.

\begin{figure}[h]
\includegraphics*[width=5.5cm]{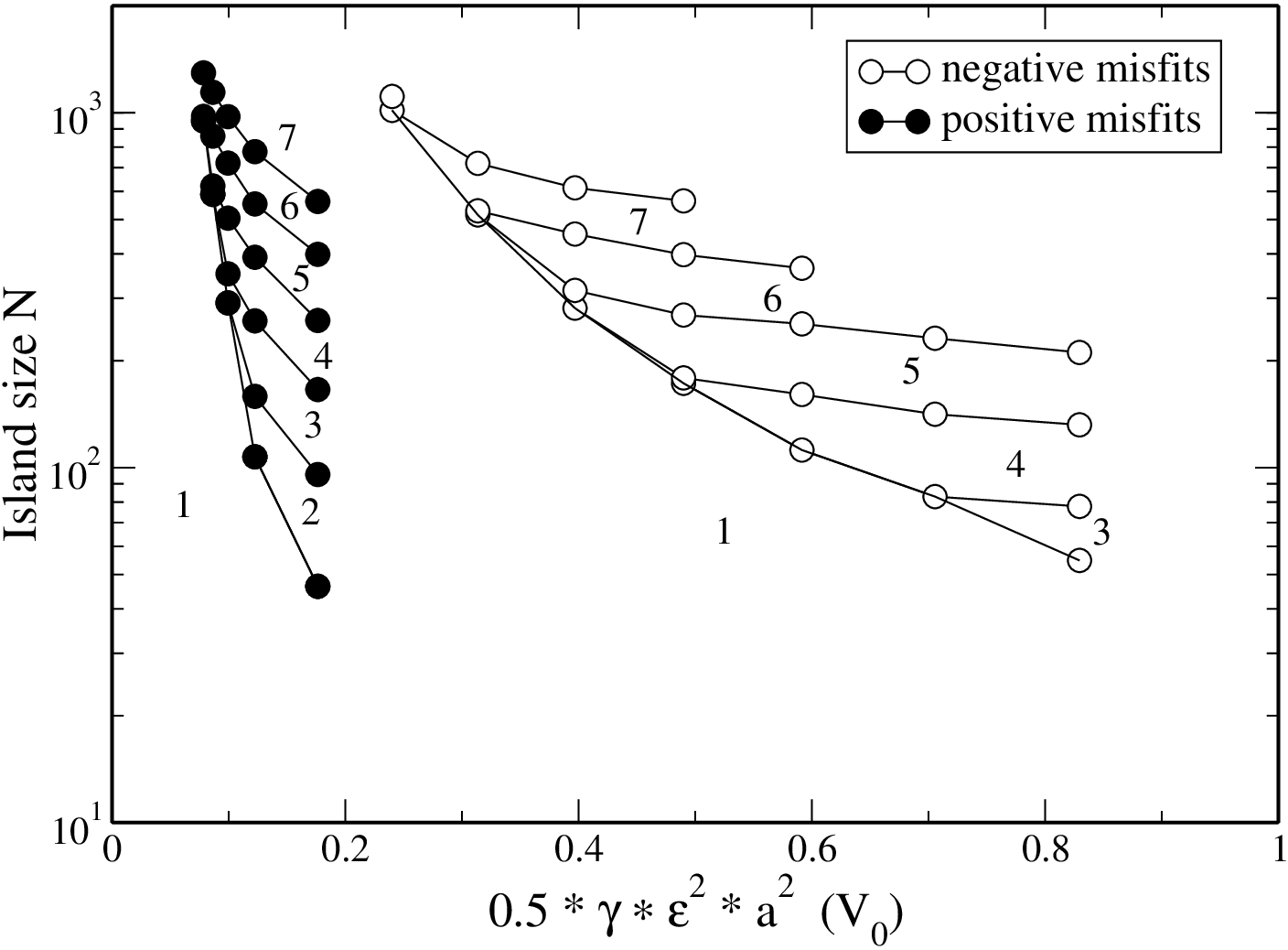}
\caption{\label{Phasedia} Phase diagram showing the stability ranges of
islands of different heights (given by the numbers in ML's) in coordinates 
of the total number of atoms and the bulk strain energy per bond 
${\cal E} = 0.5\gamma \varepsilon ^2 a^2$, for positive and negative values 
of the misfit and for $\mu = 2\nu$ potentials.}
\end{figure}

The stability regions of mono- and multilayer islands for both compressed 
and expanded overlayers are visualized in the phase diagram of 
Fig. \ref{Phasedia}, as a function of the number of atoms in the islands and 
the bulk strain energy per bond ${\cal E} = 0.5\gamma \varepsilon ^2 a^2$, 
which turns out to be the relevant scaling factor. As seen, ML islands 
are favoured at small number of atoms. For decreasing strain energies, the 
transition takes place for larger island sizes and to thicker islands. 
Expanded 
overlayers require larger strain energy to become unstable against clustering. 
Islands thinner than a certain number of layers are forbidden 
for energetic reasons.
Result are shown for the particular class of potentials with $\mu = 2\nu$;
other choices of $\mu$ and $\nu$ give points very close to those displayed.
Different potentials (e.g. Lennard-Jones) are expected to give extremely 
similar results.  

To compare with experiments, we estimate the strain energy per bond
${\cal E}$ from the force constant $\gamma$ of the overlayer material 
(in turn calculated as $\gamma = E b / 2 (1-\nu_P)$, with $E$ the 
Young modulus and $\nu_P$ the Poisson ratio). Expressing ${\cal E}$ in
units of the bond energy $V_0$ of the corresponding material, we get
0.15, 3.08 and 2.27 for the cases of Pb, Ag and Al on Si(111), or 0.05,
1.5 and 1.3 eV, respectively.
Thus, a transition for Pb/Si from 1~ML to
8 or 9~ML for islands of several thousands of atoms is expected, in
reasonable agreement with experiments showing a preferred height of 7~ML
and scarce presence of thinner 
islands~\cite{Mike1,Mike2,TTT1,Yeh,TTT2,Ozer,TTT3}. For Ag and Al, the
unrealistically high values of ${\cal E}$ (due to the failure of the
harmonic approximation for such high values of the negative misfit) predict
no absence of thin islands, as observed~\cite{TTT3,Liu} (see below).
These estimations consider (100) coherent islands while most
experiments are done on Si(111) and the islands are probably relaxed.
However, they may retain approximate validity because the strain energy 
relaxation is partially balanced by the cost of disregistry. Furthermore, 
calculating similarly for the systems In and Cd on Si (to our knowledge not 
yet studied), we obtain 0.17 and 2.65~$V_0$, respectively, thus predicting 
for In a similar behaviour as Pb (thin islands forbidden) in contrast to Cd.

\begin{figure}[h]
\includegraphics*[width=5.5cm]{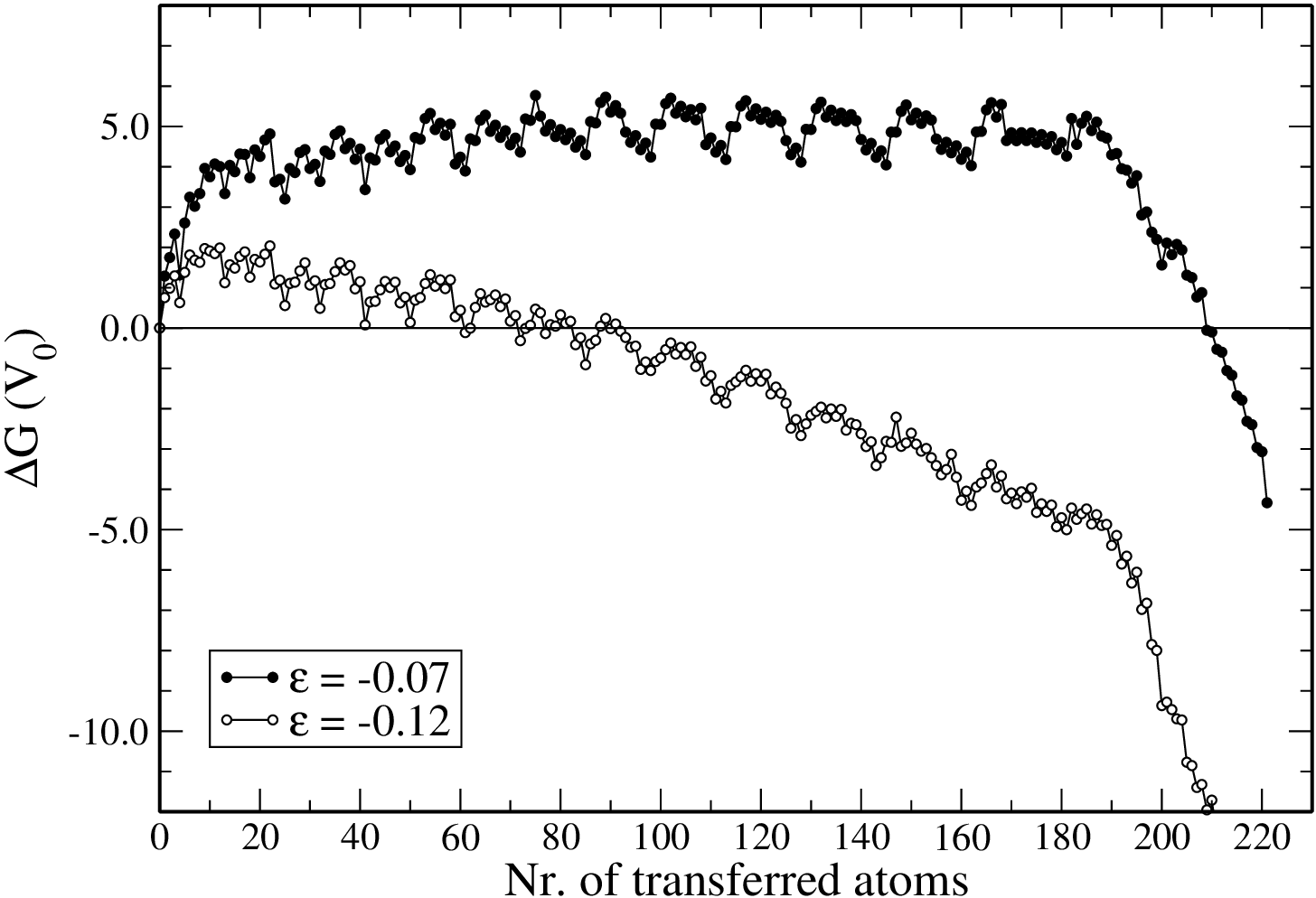}
\caption{\label{DG2613fneg-365} Transformation curves representing the energy
change in units of the bond energy $V_0$ as a function of the number of atoms
transferred to the upper level for (a) $\varepsilon = - 0.07$, and
(b) $\varepsilon = - 0.12$. The number of atoms in the initial monolayer
island ($365 \approx 19\times19$) gives a complete truncated trilayer
pyramid ($12\times12 + 11\times11 + 10\times10$) and $\mu = 2\nu = 26$.}
\end{figure}

The results above point to a possible mech\-anism of 2D-3D transformation 
different from the con\-se\-cu\-tive formation of bilayer, trilayer, etc. 
islands~\cite{Pri2,Stmar}: the direct nucleation of multilayer clusters 
whose thickness depends on the values of the force constant and the lattice 
misfit. 
We have studied this mechanism: Fig.~\ref{DG2613fneg-365} shows 
transformation curves from 1~to 3~ML islands for two negative values of 
the misfit, -0.07 and -0.12. 
Atoms are detached from the edges of the initial monolayer island and 
incorporated to the double steps of the bilayer island growing on top. 
The low-misfit curve is similar to
the layer-by-layer curve shown in Fig. 5b of Ref.~\onlinecite{Pri2}. 
The energy tends to increase 
all the way (the departure from monotonic behavior depends on the exact 
atomistic processes related to filling and depleting of atomic rows during 
the transformation process) and shows at the very 
end a sudden collapse due to the disappearance of the single and double 
steps to produce low-energy facets. On the contrary, 
the large-misfit curve shows in the beginning a nucleation behavior with a 
bilayer nucleus consisting of 22 atoms. 

We consider first transformations from mono- to bilayer islands. Two
processes accompany the island thickening: first, the dis\-tance $l$ between
the upper and lower steps decreases.
Since the step repulsion energy increases only as $l^{-2}$~\cite{Nozi},
it is negligible except very close to the end of the tranformation.
The second is the essential process, the increase of 
the total length of the steps, which in our model of square islands is
given by $\Delta L = 4(\sqrt{N_0-N_2}+ \sqrt{N_2}- \sqrt{N_0})$, with $N_0$ 
and $N_2$ being the number of atoms in the initial ML island and in 
the incomplete upper layer, repectively (note that, as a first approximation, 
there is no change in surface or interface energies, since there is no
variation of the out-of-plane coordination of the atoms, and islands and
wetting layer are composed of the same material).
This process has three energetic contributions associated: 
the first is an increase in step energy, understood as the energetic 
cost of the unsaturated bonds, due to the reduced coordination at steps.
The second is the relaxation of strain at steps and the third is the 
reduction of interlayer adhesion at steps due to the edge atoms climbing 
up in the potential valley created by their neighbors underneath. 
Only the strain relaxation favors the process of clustering~\cite{Pri2,Mul}. 
The
other two contributions oppose it, in addition to the step repulsion energy.

The sign of the misfit strongly affects the strain relaxation at steps.
For both signs and for large enough islands, the bonds are
almost completely relaxed at the edges and strained at the center.
However, for intermediate values of island's size and misfit 
(3 - 7\%)~\cite{Mil,Kor}, the region with partially relaxed bonds extends
considerably further in the case of compressed islands.
Thus, as a result of anharmonicity, strain relaxation at steps is
stronger for positive than for negative misfits of the same absolute value.
The reduced coordination at steps, i.e. the number of unsaturated bonds, 
The number of edge atoms obviously does not depend on the 
sign of the misfit, while the reduced adhesion to out-of-plane neighbours 
is expected to depend weakly, because the interaction of an edge atom with 
its neighbours in the layer underneath ne\-ces\-sa\-rily involves 
attractive and repulsive forces in both cases.
In compressed islands the strain relaxation at steps overcompensates 
the increase of the step energy (the total step length increases very fast 
in the beginning of the trans\-formation).
The balance turns for an island size which becomes very small for large
enough values of the mis\-fit. This results in nucleation-like 
behavior~\cite{Pri2}. 
In expanded islands the weaker strain relaxation at steps requires 
larger values of the misfit to overcome the remaining contributions. 

We consider next the instability of mono- against multilayer
islands at small values of the force constant of the material. 
Owing to the weak strain relaxation at steps, the strain energy per atom 
decreases very weakly with thickness.
A useful picture is to consider $n$-ML islands in coherent epitaxy as 1-ML 
islands but with $n$ times stronger bonds, i.e. 
$n$ times ``stiffer"~\cite{Merwe}. 
If the force constant of the material is small, the gain in strain energy 
upon the formation of ML islands on the first ML will be also small. 
Instead, bilayer islands will become energetically 
favoured with increasing number of atoms, i.e., the ground state will 
evolve from mono- to trilayer islands. These, being effectively
``stiffer", relax the strain energy at steps more efficiently, 
overcoming the energetic cost of the bilayer step.
For smal\-ler values of the force constant tetralayers and even thicker 
islands will become stable against monolayer islands when the total size 
increases. This behavior is found also in compressed islands, shifted to
smaller values of the force constant due to the anharmonicity of the
interatomic potential.

For both compressed and expanded overlayers, the decrease of the 
strain energy per bond leads to an increase of the critical thickness 
$X$ at which the corresponding island becomes energetically stable
against a ML island of the same size. 
A nucleation-like behavior is expected
for this 1-$X$ ML transformation. This basic result of our study, 
sum\-mar\-is\-ed in Fig. \ref{Phasedia}, implies that islands with 
insufficient thickness, i.e. with insufficient 
relaxation of strain energy at steps, will be forbidden to form for 
thermodynamic reasons. This can explain the absence of Pb islands thinner 
than 4 ML~\cite{TTT1,TTT2,Ozer,TTT3}. 

\begin{figure}[h]
\includegraphics*[width=5.5cm]{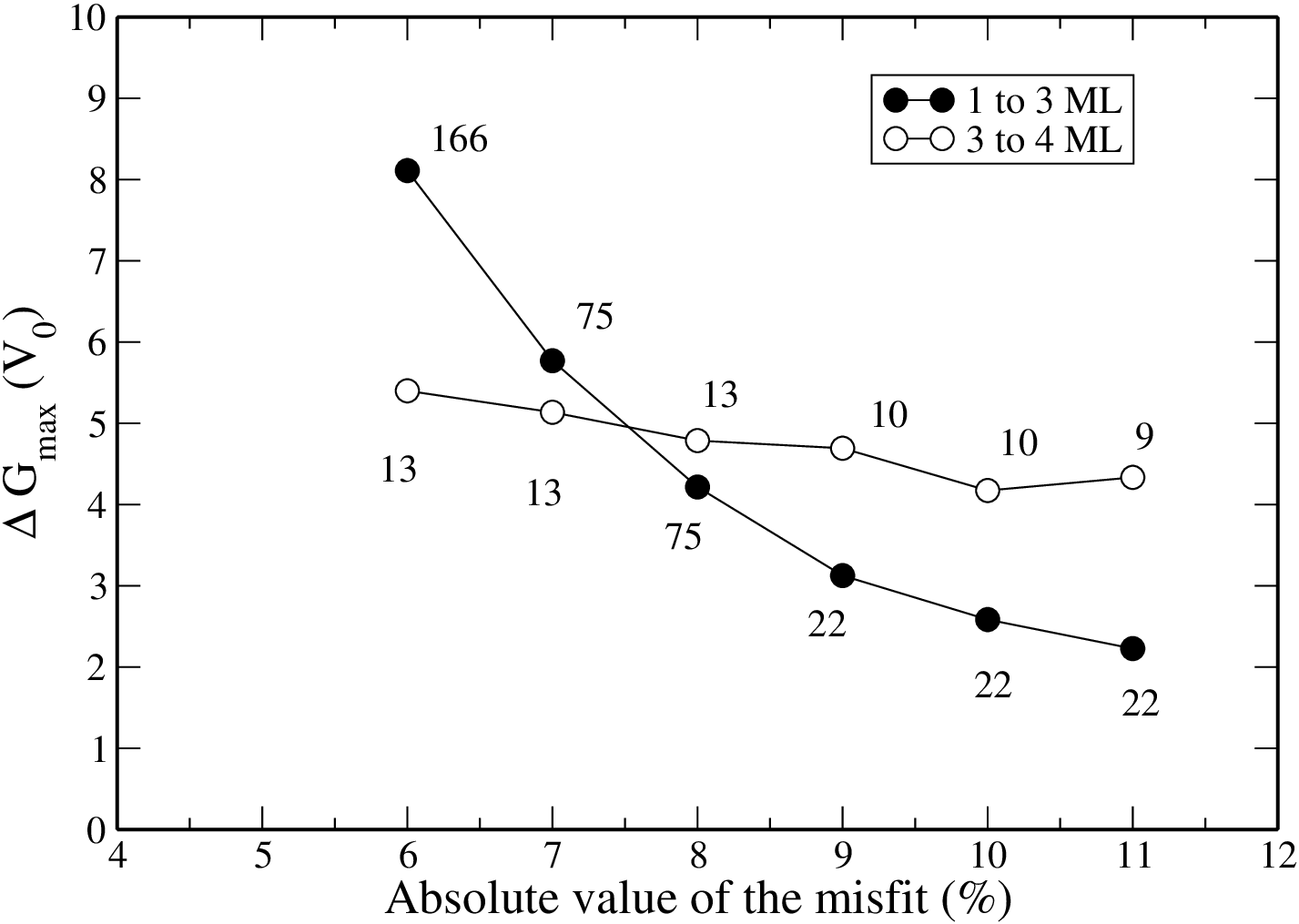}
\caption{\label{DG2vsf1-3-4} Nucleation barriers in units of $V_0$
as a function of the absolute value of the negative misfit, for mono-
to trilayer islands. The number of atoms in the critical nucleus is 
given at each point. Also shown is the barrier for monolayer nucleation 
on top of the trilayer island. The total number of atoms is 365, and 
$\mu = 2\nu =26$.}
\end{figure}

Figure \ref{DG2vsf1-3-4} shows a further important result.
The mo\-no\-to\-nic decrease of the nucleation barrier height with misfit
for the 1-3 ML transformation is characteristic of nucleation 
behaviour~\cite{Pri2}.
For small misfits, the barrier for the 3-4 ML transformation is smaller than
for 1-3 ML, implying that, if 1-3 is possible, so will be also 3-4 or even
higher. However, at larger absolute values of the mis\-fit 
the barrier for formation of a bilayer 
nuc\-leus on the first ML is significantly smaller than 
the barrier for formation of a ML nucleus on the trilayer,
so that the growth of the 4th atomic level might be kinetically 
inhibited in this case. 
A similar intersection of the curves (not shown) is found for positive 
misfits, shifted to smaller absolute values.
Thus, for large absolute values of the misfit
the strain relaxation at the ML cluster growing on the trilayer island is 
not strong enough, overcoming the increase in step energy, to result
in a small barrier height (this is essentially the same
behavior as in the 1-2 ML transformation discussed above)
Therefore, at sufficiently large misfits the growth in height can be
strongly inhibited due to kinetics. For example, a factor 2 in the barrier
height, as given by Fig.\ref{DG2vsf1-3-4} for high misfits reduces the 
nucleation rate several orders of magnitude at room temperature or below 
and 2D lateral 
growth will be preferred, as in fact observed~\cite{Mike1,Ozer,Budde}.
In addition, the side walls act as better
traps for adatoms than the flat top surface.
The lateral growth by rearrangement of mono- to multilayer
islands is strongly supported by the observation that 2-ML thick
flat-top Ag islands on Si(111) increase linearly in size with increasing
coverage whereas ML islands preserve a nearly constant size of
about 500 \AA$^2$~\cite{TTT3}, which can be interpreted as the critical
size $N_{1-2}$. A clear evidence for a transformation process is the
rearrangement of 2 to 3-ML Fe islands on Cu$_3$Au(001)
upon annealing at 400K~\cite{Canepa,Verdini}.

In conclusion, we find that materials with weak inter\-atomic bonding will 
transform from ML islands directly into thick islands by 2D multilayer 
nucleation and growth. 
Islands thinner than the crit\-ical thickness are thermodynamically forbidden 
due to the insufficient strain relaxation at steps, particularly when under
tensile stress. At sufficiently large misfits the barrier for 2D
multilayer nucleation is significantly smaller than the barrier for the
subsequent single-layer nucleation. 
The above ef\-fects due to the interplay of step energy and strain 
relax\-ation are found to be quantitatively reasonable and offer a 
plausible explanation for the absence of thin islands and the 2D 
propagation of flat-top islands in epitaxial growth of metals on 
semiconductors.

This work was supported by projects MAT2005-03011 of the Spanish MEC and
S-0505/MAT-0194 of the Comunidad de Madrid. J.E.P. also gratefully 
acknowledges financing by the programme ``Ra\-m\'on y Cajal'' and fruitful
discussions with Juan de la Figuera.

\end{document}